\documentclass[apj]{emulateapj}
\usepackage{graphicx}
\usepackage{subfigure}
\usepackage{color}
\usepackage{verbatim}
\usepackage{bm}% bold math

\newcommand*{\sgra}{Sgr~A$^\star$}

\begin{document}

%\title{A statistical study of X-ray flares in Sgr A$^\star$: solar-like self-organization criticality }
%\maketitle

\title{Statistics of X-ray flares of Sagittarius~A$^\star$: evidence for solar-like
self-organized criticality phenomenon}

\author
{Ya-Ping Li$^{1}$, Feng Yuan$^{2,1}$, Qiang Yuan$^3$, Q. Daniel Wang$^{3}$, P. F. Chen$^{4}$, Joseph Neilsen$^{5,6}$, Taotao Fang$^{1}$, Shuo Zhang$^{7}$, Jason Dexter$^{8}$ }
\affil{$^{1}$Department of Astronomy and Institute of Theoretical Physics and Astrophysics,
Xiamen University, \\ Xiamen, Fujian 361005, China; leeyp2009@gmail.com}
\affil{$^{2}$Shanghai Astronomical Observatory, Chinese Academy of Sciences, 80
Nandan Road, Shanghai 200030, China; fyuan@shao.ac.cn}
\affil{$^{3}$Department of Astronomy, University of Massachusetts, Amherst, MA 01003, USA; yuanq@umass.edu; wqd@astro.umass.edu}
\affil{$^{4}$School of Astronomy and Space Science, Nanjing University, Nanjing 210023,
China}
\affil{$^{5}$ Boston University Department of Astronomy, Boston, MA 02215, USA}
\affil{$^{6}$ MIT Kavli Institute for Astrophysics and Space Research, Cambridge, MA 02139, USA}
\affil{$^{7}$ Columbia Astrophysics Laboratory, Columbia University, New York, NY 10027, USA}
\affil{$^{8}$Max Planck Institute for Extraterrestrial Physics, P.O. Box 1312, Giessenbachstr., D-85741 Garching, Germany}
%\affil{$^{6}$Department of Astronomy, Hearst Field Annex, University of California, Berkeley, CA 94720-3411, %USA}

\begin{abstract}
X-ray flares have routinely been observed from the supermassive black hole, Sagittarius~A$^\star$ (\sgra), at our Galactic center. The nature of these flares remains largely unclear, despite of many theoretical models. In this paper, we study the statistical properties of the \sgra\ X-ray flares, by fitting the count rate (CR) distribution and the structure function (SF) of the light curve with a Markov Chain Monte Carlo (MCMC) method. With the 3 million second \textit{Chandra} observations accumulated in the \sgra\ X-ray Visionary Project, we construct the theoretical light curves through Monte Carlo simulations. We find that the $2-8$ keV X-ray light curve can be decomposed into a quiescent component with a constant count rate of $
\sim6\times10^{-3}~$count~s$^{-1}$ and a flare component with a power-law fluence distribution $dN/dE\propto E^{-\alpha_{\rm E}}$ with $\alpha_{\rm E}=1.65\pm0.17$. The duration-fluence correlation can also be modelled as a power-law $T\propto E^{\alpha_{\rm ET}}$ with $\alpha_{\rm ET} < 0.55$ ($95\%$ confidence). These statistical properties are consistent with the theoretical prediction of the self-organized criticality (SOC) system with the spatial dimension $S = 3$. We suggest that the X-ray flares represent plasmoid ejections driven by magnetic reconnection (similar to solar flares) in the accretion flow onto the black hole.
\end{abstract}

\section{introduction}

Sagittarius~A$^\star$ (\sgra) at the center of the Milky Way is an excellent laboratory for studying the accretion and ejection of matter by supermassive black holes (SMBHs).
There have been quite a number of observational and theoretical
studies of \sgra\ (see reviews by \citealt{Genzel10} and \citealt{Yuan14}).
The bolometric luminosity of \sgra\ is $L_{\rm bol}\sim 10^{-9}L_{\rm Edd}$ (where $L_{\rm Edd}$ is the Eddington luminosity), which
is five orders of magnitude lower than that predicted by a standard thin
disk accretion at the Bondi accretion rate \citep{Baganoff03}. We now understand
that an advection-dominated accretion flow scenario works for \sgra, and that the
low luminosity is due to the combination of the low radiative efficiency and
the mass loss via outflow \citep{Yuan14,Yuan03,Yuan12,Narayan12,Li13,Wang13}.
\sgra\ is usually in a quiescent state, and occasionally shows rapid
flares (on time scales $\sim1~$hour), most significantly in X-ray \citep{Baganoff01} and near-infrared (NIR; \citealt{Genzel03,Ghez04}). The flare rate is roughly once
per day in X-ray and more frequently in NIR.

Theoretical interpretations of the flares include the electron acceleration by, e.g.,
shocks, magnetic reconnection, or turbulence, produced in either an
accretion flow (e.g., \citealt{Yuan04,Eckart06,Yusef09,DoddsEden09,DoddsEden10,Chan15}) or in
an assumed jet (e.g., \citealt{Markoff01}).
The other scenarios in terms of the flaring cause are transient features in the accretion flow, such as accretion instability \citep{Tagger06}, orbiting hot spot \citep{Broderick05}, expanding plasma blob \citep{Yusef06,Eckart06,Yusef09,Trap11}, and tidal disruption of asteroids \citep{Zubovas12}.
However, the nature of the flares is still under debate.

Before 2012, only about a dozen of X-ray flares were detected (e.g.,   \citealt{Baganoff01,Porquet03,Belanger05,Eckart06,Aharonian08,Marrone08,Porquet08,Yusef08,Degenaar13}). Thanks to the {\it Chandra} X-ray Visionary Project on \sgra\ (hereafter XVP\footnote{http://www.sgra-star.com/}) in 2012, a total of 39 flares have been added  \citep{Neilsen13}. This substantially increased sample size of the flares has enabled the statistical study of the properties. \citet{Neilsen13} first analyzed the distributions of the fluences, durations, peak count rates and luminosities of the XVP detected flares. \citet{Neilsen15} further investigated the flux distribution of the XVP X-ray light curves taking into account the Possion fluctuations of the quiescent emission and the flare flux statistics.

In this work we advance the statistical analysis of the XVP light curve by simultaneously using its flux distribution and structure function (SF). We generate X-ray light curves through Monte Carlo realizations of both the quiescent and flare contributions. The realization of flares assumes power law models for their fluence function and  width-fluence dependence.  The parameters of these power laws are in turn constrained by comparing the simulated data with the observed one, using the Markov Chain Monte Carlo (MCMC) approach.

The major motivation of the statistical analysis is to determine whether the SOC model can explain the \sgra\  X-ray flares, and if so to probe the dimensionality of the process leading to the flares. We compare our constrained power law indexes (i.e., the power law indexes for the fluence distribution and the duration-fluence correlation) with the expectation of the self-organized criticality (SOC) theory, which describes a class of dynamical system with nonlinear energy dissipation that is slowly and continuously driven toward a critical value of an instability threshold (\citealt{Katz86,BTW,Aschwanden11}; see the Appendix for a brief introduction to SOC and the relevant statistical method to be used in the present work). The energy dissipation driven by magnetic reconnection is believed to be in an SOC system, such as solar flares \citep{Lu91}, possibly the X-ray flares of $\gamma$-ray bursts (GRBs) and black holes of various scales \citep{WD13,Wang15}. An SOC system experiences scale-free power-law distributions of various event parameters, such as the total energy (or equivalently the fluence), the peak energy dissipation rate, the time duration, and the flux (or equivalently the CR) of events.
These statistical properties are related to the geometric dimension of the system that  drives the events, and can thus be used to diagnose their physical nature \citep{Aschwanden12,Aschwanden14}. While the statistical analysis of solar flares suggests a spatial dimension $S=3$, consistent with the complex magnetic structure observed in active regions on the Sun \citep[e.g.,][]{Priest00}, the X-ray flares of GRBs favour $S=1$ \citep{WD13}. Thus it is argued that the X-ray flares occur in the jets of GRBs where the magnetic field configuration tends to be poloidal and thus one-dimensional \citep{WD13}. \citet{Wang15} further claim $S=3$ from an analysis of the X-ray flares of \sgra, Swift J1644+57, and M 87. However, this analysis is very crude in the sense that the correlation among different bins of the cumulative distribution are not properly addressed, that artificial cuts of the fitting range of the data are
adopted, and that the used flare sample is strongly biased and incomplete at the detection limit \citep{Wang04}. All these issues are handled in the present work.

The rest of this paper is organized as follows.
The observational data and the statistical method adopted are described
in \S2. We present the fitting results in \S3, and discuss the physical
implication of the results in \S4. Finally we conclude this work in \S5.

\section{data and methodology}

\subsection{Observations}
We use the data of the 2012 \textit{Chandra} XVP campaign, which consists
of 38 \textit{Chandra} ACIS-S/HETGS observations of \sgra\ with the total exposure
time of 3 million second (Ms) between February 6 and October 29 in 2012. The $2-8$ keV
light curve is extracted in 300 s bins including both photons
of the $0^{\rm th}$ order and the $\pm1^{\rm st}$ diffraction orders.
The $0^{\rm th}$ order events are extracted from a circle region with
a radius of $1''.25$, while the $\pm1^{\rm st}$ order events are extracted
from a $2''.5$ rectangular region \citep{Neilsen13}. For more details
of the XVP campaign and the data reduction, we refer the readers to
\citet{Neilsen13}.

\subsection{Synthetic Light Curve}
The X-ray light curve can be decomposed into the quiescent and
flare components. The quiescent emission is assumed to be steady with a count rate (CR) $r$.
We model the flare component as a sum of Gaussian
functions\footnote{This modelling is used as the 1$^{\rm st}$-order approximation to account
for the correlation among photons from  individual flares, although some bright ones
do show significant asymmetric shapes \citep{Baganoff01,Nowak12}, and/or substructures \citep{Barriere14}}. Then the model light curve is
\begin{equation}\label{lc:def}
R\left(t\right) = \sum_{i=1}^{[\kappa]}\frac{E_i}{\sqrt{2\pi}\sigma_i}
\exp\left[-\frac{1}{2}\left(\frac{t-\mu_i}{\sigma_i}\right)^2\right] + r,
\end{equation}
where $E_i$, $\sigma_i$, and $\mu_i$ are the fluence, width and peak location
of the $i$th flare, $[\kappa]$ is the integer part of $\kappa$ and it presents the total number of flares. We use a Monte Carlo method to generate the model light curve, which accounts for
the Possion fluctuations of photon counting. The differential flare fluence function is
assumed to be a power-law form,
\begin{equation}\label{pl:def}
dN/dE \propto \left \{
\begin{array}{l@{\quad \quad}l}
E^{-\alpha_{\rm E}}, &  E_{\rm{min}}\leq E \leq E_{\rm{max}}\\
0, & \rm{otherwise} ~~
\end{array} \right.,
\end{equation}
where the lower limit of the fluence $E_{\rm{min}}=1~\rm{cts}$ is well below
the detectable flare fluence in \citet{Neilsen13}, and the upper limit
$E_{\rm{max}}=1000~\rm{cts}$ is slightly larger than that of the brightest
flare \citep{Nowak12}. Given $\alpha_{\rm E}$ and $\kappa$, the fluences of
the model flares can be easily sampled.

As shown in \citet{Neilsen13}, there is a clear correlation between the
flare fluence $E$ and duration $T$ (defined as 4 times of the Gaussian width
$\sigma$). We characterize this correlation with a log-normal distribution
$N(\log\sigma_i, \sigma_{\rm ET})$, where  $\sigma_{\rm ET}$ is the
Gaussian dispersion and  a power-law relation between $\sigma_i$ and $E_i$ as
\begin{equation}\label{ETrel:def}
\sigma_i=A¡«\left(\frac{E_{i}}{100}\right)^{\alpha_{\small{\rm ET}}}£¬
\end{equation}
%%$$\sigma_i=A¡«\left(\frac{E_{i}}{100}\right)^{\alpha_{\small{\rm ET}}}£¬$$
%%\log \sigma_i=\log A +  {\alpha_{\small{\rm ET}}}\log (E_{i}/100)£¬
where $A$ is the normalization constant and $\alpha_{\rm ET}$ is the
power-law index\footnote{In this work we use $\alpha_{\rm X}$ to represent
the power-law index of the distribution of variable $X$ (specifically
$dN/dX\propto X^{-\alpha_{\rm X}}$), and $\alpha_{\rm XY}$ to represent
the correlation power-law index between $X$ and $Y$ (specifically
$Y\propto X^{\alpha_{\rm XY}}$).}. The intrinsic scattering of the flare
widths around the power-law relation is estimated to be $\sigma_{\rm ET}=0.25$ for the detected sample
(Yuan et al. 2015, in preparation).  The peak time $\mu_i$ of each flare
is generated randomly in the observational periods. The discrete light curve is then generated in 300-s bins, resulting in $M=9964$ bins for the total exposure of the data. The average count rates are calculated in each bin of the simulated light curve. In total we will have five parameters to
describe the synthetic light curve, namely the Poisson background $r$, the power-law index of the fluence distribution $\alpha_{\rm E}$, the total flare number $\kappa$,
the duration normalization $A$, and the fluence-duration correlation slope $\alpha_{\rm ET}$.

Finally we correct the pileup effect of the simulated light curve by assuming the $1^{\rm st}$-order count rates a constant fraction, $52\%$, of the incident rate (\citealt{Nowak12}, see also \citealt{Neilsen13,Neilsen15}).
For all the simulated CRs, the pileup correction is less than $20\%$.

\begin{figure*}[!htbp]
\begin{center}
\includegraphics[width=0.95\textwidth]{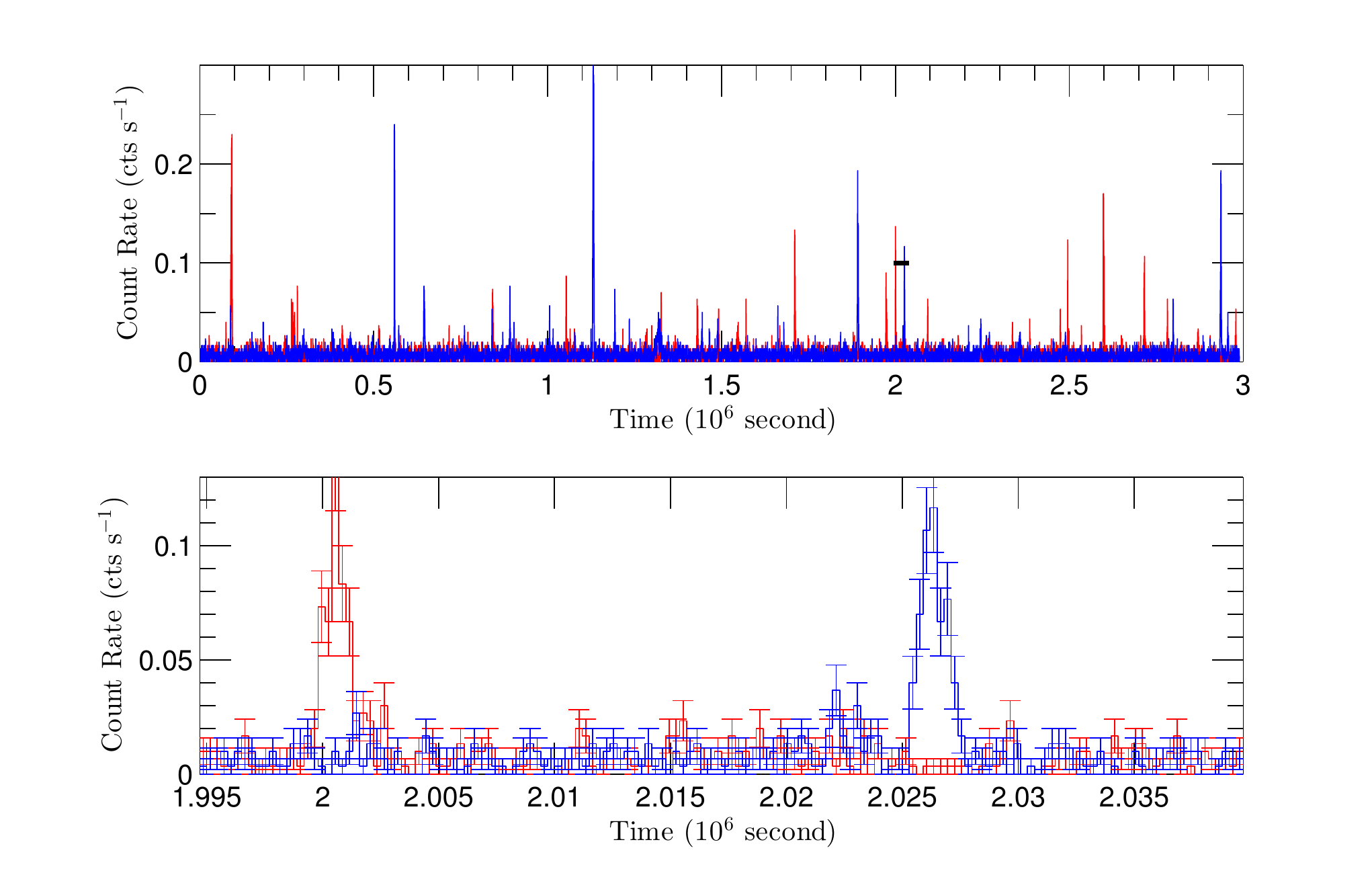}
\end{center}
\caption{\small Upper panel: The light curve (with the observing gaps
removed) from the XVP observations (red), compared with a representative
one from the simulations (blue) using the best-fit parameters of model I
shown in Table \ref{tab:para} (see below). Lower panel: Zoomed-in light curve
of a time window indicated by the short horizontal bar in
the upper panel. }\label{fig:lc:sf}
\end{figure*}

Fig.~\ref{fig:lc:sf} shows a representative simulated light curve (blue), compared
with the observational one (red) with the observing gaps removed.

\subsection{Statistical Comparison}
The synthetic light curve cannot be directly compared to the observational
one due to the lack of bin-to-bin correspondence between the two.
We need a statistical way to make the comparison. A ``first'' order
statistics is the CR distribution, which reveals the overall magnitude of
the variabilities. However, as mentioned above (also noted in
\citealt{Neilsen15}), the CR distribution contains no information about
the flare flux correlation among adjacent bins. This correlation can be
independently  accounted for by the use of the auto-correlation function,
or equivalently the SF for a stationary process \citep{Simonetti85,Emmanoulopoulos10}.
We jointly fit the CR distribution and the SF to determine the model
parameters.

\subsubsection{Count Rate Distribution}\label{app:stat}
With the 300-s bins both for the simulated and observed data, the binning results in $M=9964$ data points for the total exposure window. We construct the CR distribution of the data logarithmically and
follow Knuth~(2006; see also \citealt{Witzel12}) to determine the optimal
bin number of the histogram. Considering the histogram as a piecewise-constant model, the
relative logarithmic posterior probability (RLPP) for the bin number $m$, given $M$
data points, is
\begin{eqnarray}\label{RLPP}
\nonumber
  {\rm RLPP}(m)&=&M\log m+ \log\Gamma\left(\frac{m}{2}\right)- m\log\Gamma\left(\frac{1}{2}\right)\\
  &-&\log\Gamma(M+\frac{m}{2}) + \sum^{m}_{l=1}\log\Gamma\left(n_l+\frac{1}{2}\right),
\end{eqnarray}
where $n_l$ is the histogram value of the $l$th bin, and $\Gamma(x)$ is
the Gamma function. Then the optimal number of bins $\bar{m}$ can be derived
through maximizing the above $\rm{RLPP}$. For the observational light curve
of \sgra, we find $m\sim14$.

With the above binning, we can now construct $\chi^2$ statistics for the CR distribution analysis:
\begin{equation}\label{chi:def}
\chi^2_{\rm CR}=(\textbf{n}_{\rm obs}-\textbf{n}_{\rm sim})^T\textbf{C}^{-1}
(\textbf{n}_{\rm obs}-\textbf{n}_{\rm sim}),
\end{equation}
where the symbol $^T$ stands for the transpose of the matrix (vector),
$\textbf{n}_{\rm obs}$ and $\textbf{n}_{\rm sim}$ are the vectors of the
CR histogram values of the observed and simulated data, respectively.
The covariance matrix $\textbf{C}$ is calculated according to $N$ Monte
Carlo realizations
\begin{eqnarray}\label{Cov:def}
C(i,j)=\frac{1}{N-1}\sum_{k=1}^{N}\left(n_i^k-\bar{n}_i\right)
\left(n_j^k-\bar{n}_j\right),
\end{eqnarray}
where $\bar{n}_i$ is the mean value of the $i$th bin of $N$ realizations.
The covariance matrix is involved here to account for the error correlations among different bins due to the correlated flare CRs (\citealt{Brockwell02}, and see discussions of the error estimation in \citealt{Norberg09}). If there is no correlation between two histogram bins,
$C(i,j)=\delta_{ij}\sigma_i^2$, Eq. (\ref{chi:def}) is reduced to the
standard definition of $\chi^2$. The exact value of $N$,
as long as it is sufficiently large, has little effect on the calculation
of the covariance matrix. In this work we adopt $N=500$.

\subsubsection{Structure Function}\label{sect:sf:def}
Following \citet{Emmanoulopoulos10} we define the normalized SF as
\begin{equation}\label{sf:def}
V(\tau) = \frac{\left\langle\left[R(t+\tau)-R(t)\right]^{2}\right\rangle}
{S^2},
\end{equation}
where $R(t)$ is the CR as a function of time $t$, $\tau$ is the time
lag, $\left\langle\right\rangle$ represents the average over the whole time
range of the light curve, and $S^2$ is the variance of the CR
\begin{equation}
S^2 = \left\langle\left[R(t)-\overline{R(t)}\right]^{2}\right\rangle.
\end{equation}
This normalized SF is equivalent to the auto-correlation
function as long as the time series is stationary, i.e., $\overline{R(t)}=
\overline{R(t+\tau)}=\textrm{const.}$ \citep{Emmanoulopoulos10}. The
SF is especially suitable for the analysis of the data that are unequally
sampled with large observational gaps, as is the case here.
%A more comprehensive study on the pros and cons of
%the SF method can be found in \citet{Emmanoulopoulos10}.

Similar to the CR distribution, we build a $\chi^2$ statistics to compare the
observed and simulated SFs quantitatively. Apart from the correlation among various SFs
which needs to be taken into account more seriously compared with the CR distribution,
another issue for the standard definition of the $\chi^2$ is that the random variable
(the SF here) needs to be Gaussian. As shown in \citet{Emmanoulopoulos10},
the logarithm of the SF instead of the SF itself is approximately Gaussian\footnote{Although \citet{Emmanoulopoulos10} show that the logarithm of the SF is only approximately Gaussian below a false break time scale related with its power spectrum, our Monte Carlo simulations suggest that $\log V(\tau)$ is indeed distributed roughly normally for the time scales considered here in spite of its ignorance of the location of the false break. }. The $\chi^2$ statistics for the SF is thus defined as
\begin{equation}
\chi^2_{\rm SF}=(\log\textbf{V}_{\rm obs}-\log\textbf{V}_{\rm sim})^T\textbf{C}_{\rm SF}^{-1}(\log\textbf{V}_{\rm obs}-\log\textbf{V}_{\rm sim}),
\end{equation}
where the covariance matrix $\textbf{C}_{\rm SF}$ can be calculated the
same way as Eq. (\ref{Cov:def}). The number of the SF bins for the fitting procedure is chosen to be the same as that of the CR bins, i.e., $m=14$. Note that the SF becomes a featureless plateau when the time lag is larger than the largest time scale of the flares. Correspondingly, the SF for the 14 bins are calculated between 300~s and $3\times10^4$~s.\footnote{Firstly, the SFs for the larger time delay $\tau$ will show remarkable fluctuations, and thus can be impossibly used for the fitting. Secondly, we calculate the SFs with the observing gaps removed, which could introduce an artificial dip in the SFs when $\tau\gtrsim8\times10^4$~s, corresponding to the typical time scale of the observing windows for the XVP data. However, the observing gaps show no significant effect on the SFs with $\tau\lesssim3\times10^4$~s, the fitting range considered here.} By minimizing $\chi^2= \chi^2_{\rm CR} + \chi^2_{\rm SF}$ , we test our model and constrain the model parameters.

\subsubsection{Model Fitting}
We minimize $\chi^2= \chi^2_{\rm CR} + \chi^2_{\rm SF}$, using the MCMC method. It maps out the full posterior probability distributions and correlations of the model
parameters. The MCMC code is adapted from the public CosmoMC code
\citep{Lewis02}. The Metropolis-Hastings algorithm \citep{Metropolis53,
Hastings70}, a propose-and-accept process in which the acceptance or
rejection of a proposed point in the parameter space depends on the
probability ratio between this point and the previous one, is adopted
to generate the Markov chains.

\section{Results}

\subsection{Joint Fitting} \label{sect:sim}
\begin{figure}[!htbp]
\begin{center}
\includegraphics[width=0.45\textwidth]{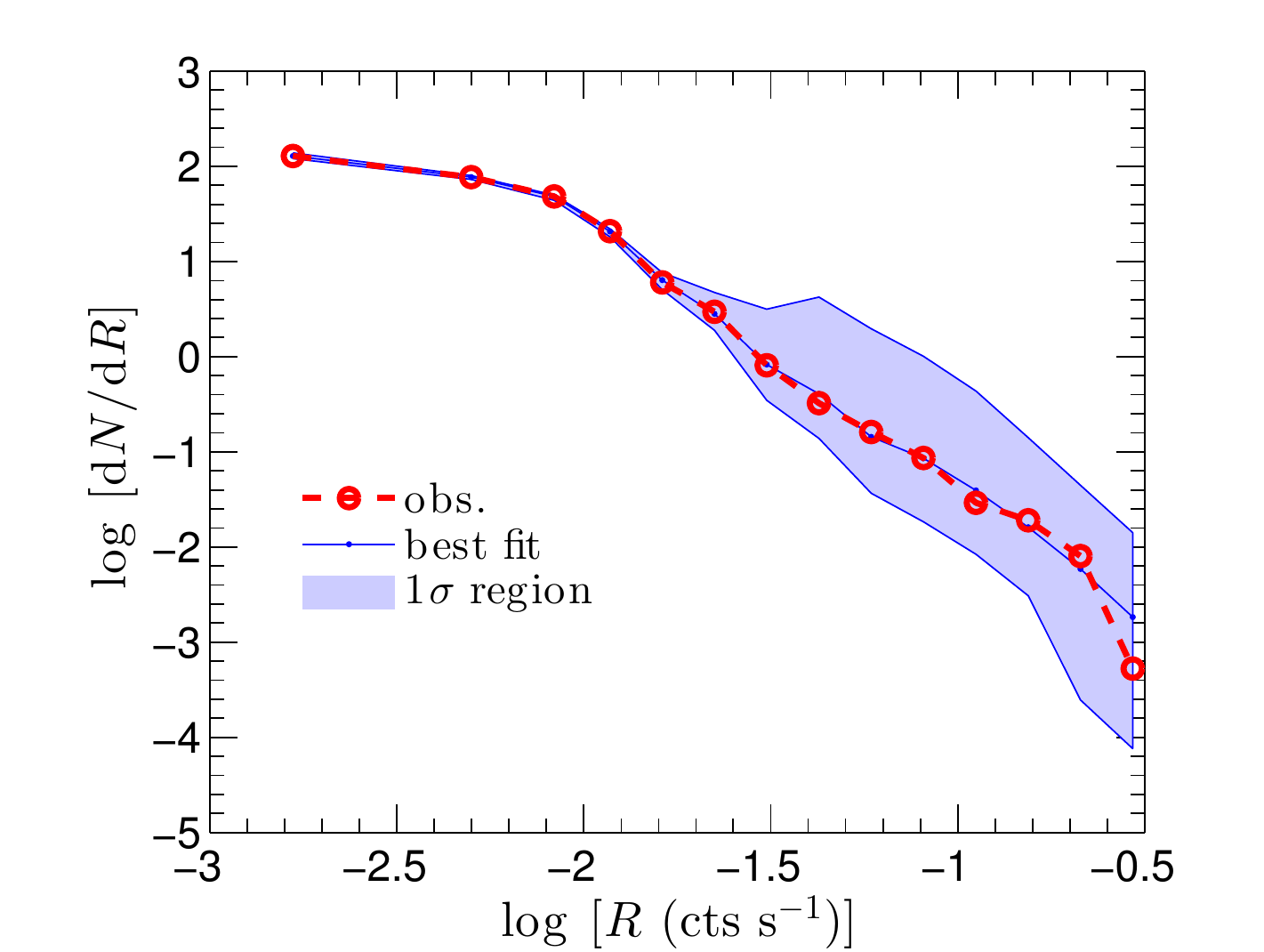}
\includegraphics[width=0.45\textwidth]{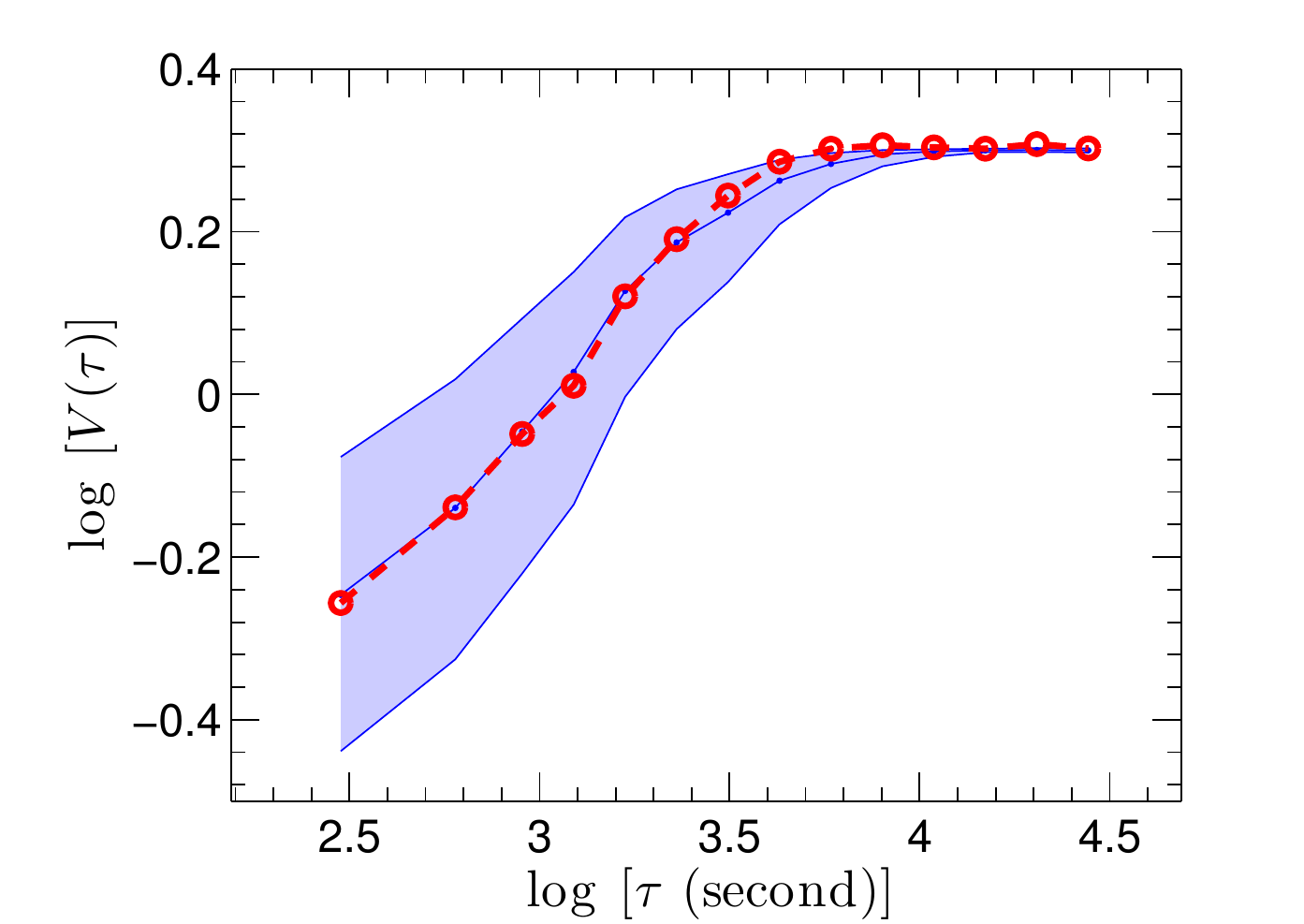}
\end{center}
\caption{\small Upper panel: The CR distribution of the X-ray light curves of \sgra. The red circles are for the observed light curve, and blue dots are the best-fit model results. The bands show the fitting $1\sigma$ ranges of the CR distribution for the best-fit model according to the MCMC chains. Lower panel: The SF for the observed data and the simulations for the same parameter set.
 }\label{fig:nsf_cr}
\end{figure}

The parameter $\sigma_{\rm ET}$ in the fittings is chosen to be 0.25 according to Yuan et al. (2015). Fig.~\ref{fig:nsf_cr} shows the best-fit results and $1\sigma$ bands of the CR (upper panel) and the SF (lower panel), compared with the observations. The CR distribution exhibits an exponential
distribution, which reflects the Possion background, and a power-law tail
which represents the flares. The SF increases with $\tau$, and reaches a plateau of $\tau\sim5\times10^3$ s, which corresponds to the largest duration of the flares. The profile with a rising trend clearly
shows the time scale of the flares of the light curve. When the time lag $\tau$ is much longer than the largest time scale of the variabilities, the normalized SF asymptotically approaches  2, i.e., the plateau in Fig. \ref{fig:nsf_cr}. It should be noted that the SF can be biased when applied to irregularly sampled light curves. In particular, the SF can show spurious breaks at $\Delta t \sim 10^{-1} \Delta T$, where $\Delta T$ is the total exposure time of the light curve \citep{Emmanoulopoulos10}. However, the break seen here is at a much shorter time scale ($\sim10^{-3}\Delta T$), and consistent with break time scales seen previously in both the NIR \citep{Meyer09, Witzel12} and sub-millimeter \citep{Dexter14}. Both of them suggest that the turnover time is likely an intrinsic characteristic time scale for the X-ray flares.

\begin{figure*}[!htbp]
\begin{center}
\includegraphics[width=0.95\textwidth]{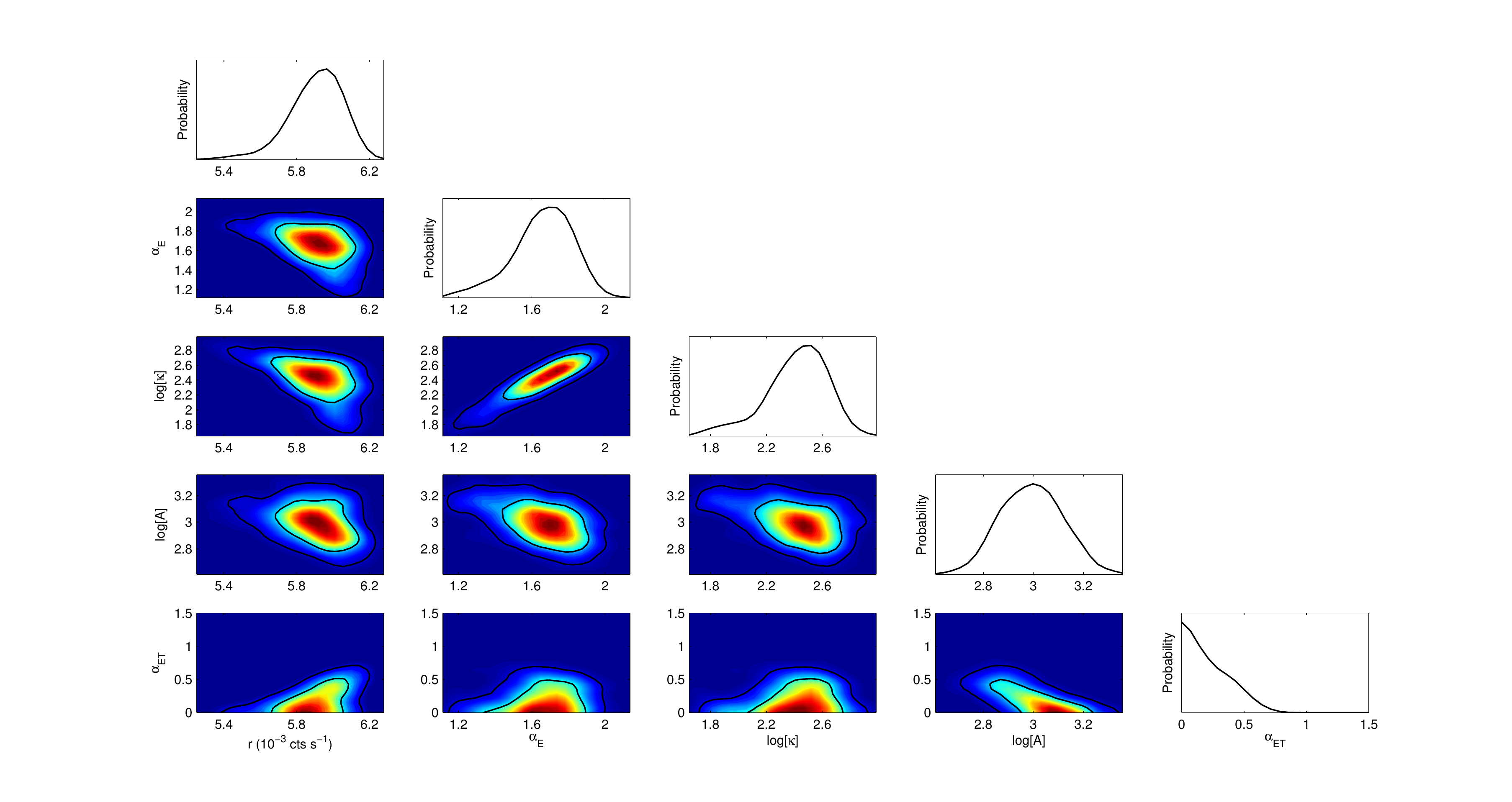}
\end{center}
\caption{\small 1D and 2D probability distributions of the fitting parameters. The contours in the 2D plots are for confidence
levels $68\%$ and $95\%$ from inside to outside, respectively.
}\label{fig:dist:nsf_cr}
\end{figure*}

Fig. \ref{fig:dist:nsf_cr} shows the one and two dimensional (1D and 2D)
probability distributions of the parameters. The mean values and the $1\sigma$ uncertainties of the parameters are listed in Table \ref{tab:para}. The fitting results suggest a fluence distribution power-law index $\alpha_{\rm E}=1.65\pm0.17$, and the less constrained fluence-duration correlation index  $\alpha_{\rm ET}<0.55~$ $(95\%~$ confidence limit). The correlations between some of the parameters can be clearly seen in Fig. \ref{fig:dist:nsf_cr}. There is a strong correlation between the total number of flares $\kappa$ and the fluence distribution index $\alpha_{\rm E}$. It is easy to understand that a harder fluence distribution will naturally correspond to a smaller number of flares in order not to over-produce the number of photons. Also the anti-correlation
between $\alpha_{\rm E}$ (or $\kappa$) and $r$ is again due to the constraint of the total number of photon counts.

\begin{table}[!htbp]
\centering
\caption{Best-fit model parameters with the MCMC method}
\label{tab:para}
\resizebox{0.45\textwidth}{!}{%
\begin{tabular}{ccccccc}
\hline\hline
    %\hline
  $r\ (10^{-3}~\rm{cts~s^{-1}})$ 	& $\alpha_{\rm E}$ 	& $\log(\kappa)$ & $\log(A)$ & $\alpha_{\rm ET}$&$\chi^2_{\nu}$ \\
  \\
 %\hline
 %\\
 $5.90\pm0.14$& $1.65\pm0.17$ & $2.41\pm0.22$ & $2.99\pm{0.12}$ & $<0.55$ & $0.9$\\
 %\\
  \hline\hline
\end{tabular}
}
\tablecomments{\small The errors are all $1\sigma$ (or $68\%$) confidence limit except for $\alpha_{\rm ET}$ (for the one tail $95\%$ confidence).\\
}
\end{table}

\begin{figure}[!htbp]
\begin{center}
\includegraphics[width=0.45\textwidth]{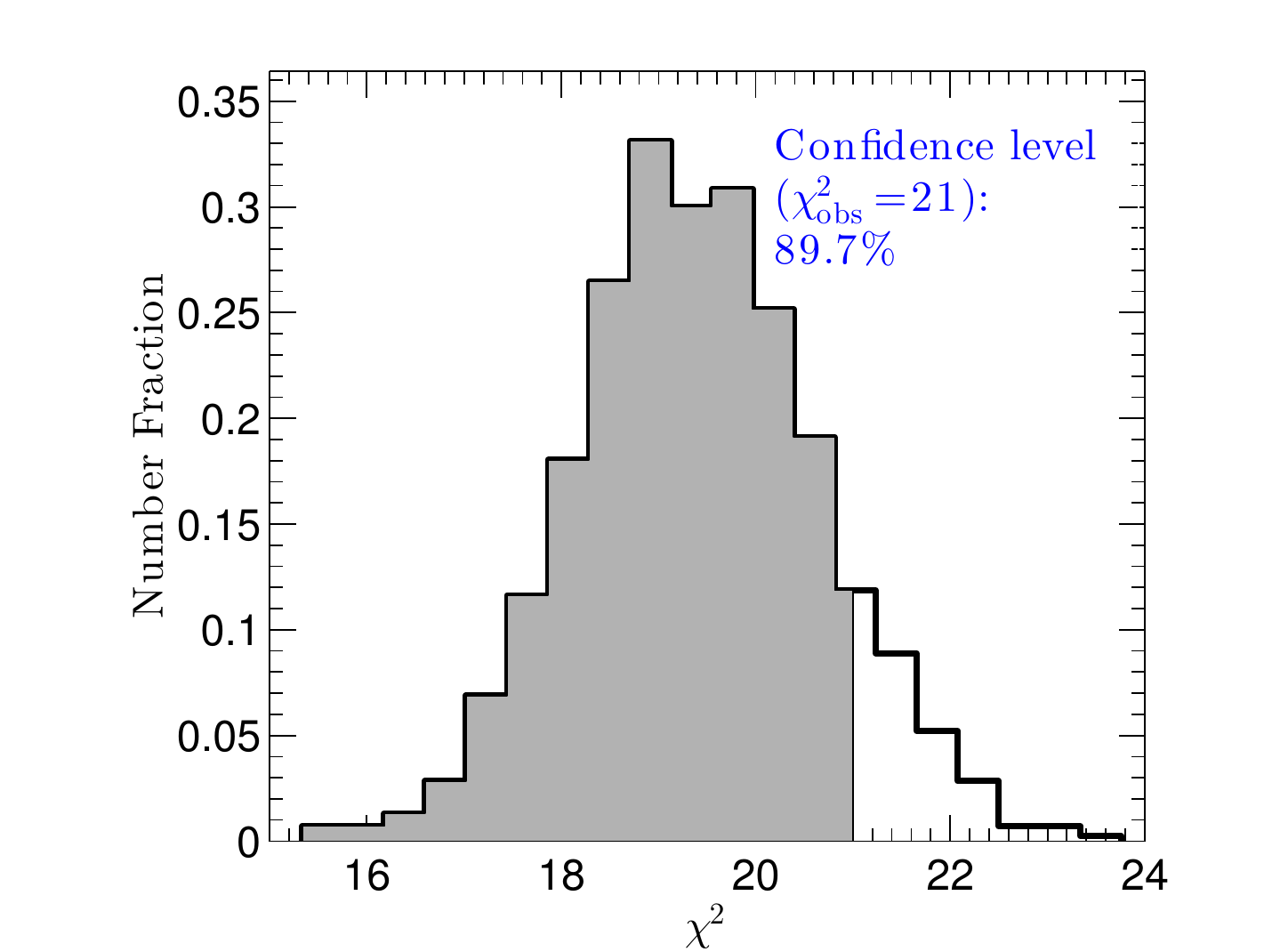}
\end{center}
\caption{\small The distribution of $\chi^2$ values of 2000 realizations
of the light curves with the best-fit model parameters. The shaded region corresponds to a percentile of $89.7\%$ for $\chi^2<\chi^2_{\rm obs}$.
}\label{fig:gof}
\end{figure}

To judge the ``goodness of fit'' of the best-fit model, we use the bootstrapping method to estimate the confidence level. Based on the best-fit model parameter set as shown in Table \ref{tab:para} we generate 2000 realizations of the light curves and calculate the $\chi^2$ values for each realization using the same method described in \S2.3. The distribution of the $\chi^2$ values is shown in Fig. \ref{fig:gof}. The number fraction of realizations with $\chi^2$ values smaller than that of the observational one, $\chi^2_{\rm obs}=21$, is estimated to be $89.7\%$, indicating that the observed light curve is well characterized by the best-fit model.

%\begin{figure}[htb]
%\begin{center}
%\includegraphics[width=0.45\textwidth]{Fig/ET.eps}
%%\includegraphics[width=0.45\textwidth]{ET_II.eps}
%\end{center}
%\caption{\small The fluence versus duration distribution of the detected
%flares (red; \citealt{Neilsen13}) and simulated ones (blue). }\label{fig:ET}
%\end{figure}

Our preliminary re-analysis of the light curve indicates that the existing sample of  flares may be biased against the detection of long duration ones. This bias can significantly affect the measurement of $\sigma_{\rm ET}$ and potentially other parameters as well. We have thus tested this latter sensitivity by fixing  $\sigma_{\rm ET}$ to different values and find that the estimates of other parameters are not significantly altered (well within their uncertainties). The analysis, however, may be sensitive to the assumed specific shape of individual flares. We will address these potential higher-order complications in a follow-up paper.

\subsection{Comparison with previous works}

\cite{Neilsen13} studied the statistical properties of the fluences, peak
rates, durations, and luminosities, of the 39 flares detected during the
2012 \textit{Chandra} XVP. It is found that the distributions of the
durations and luminosities can be well described by power-law functions,
with indices $\alpha_{\rm T}=0.9\pm0.2$ and $\alpha_{\rm L}=1.9^{+0.4}_{-0.3}$,
respectively. The power-law fittings to the fluence and peak rate distributions
give $\alpha_{\rm E}=1.5\pm0.2$ and $\alpha_{\rm P}=1.9^{+0.5}_{-0.4}$.
However, the fittings can be significantly improved by assuming cutoff power-law
functions \citep{Neilsen13}. Our result of $\alpha_{\rm E}$ are consistent
with that derived in \citet{Neilsen13}, even though only the detected
sample was employed in the latter. Based on the Gaussian flare profile
assumption, the slopes of the duration and peak rate distributions can be
derived according to our fitting results\footnote{For the duration
distribution, since
$T\propto E^{\alpha_{\rm ET}}$, we have $\frac{dN}{dT}=\frac{dN}{dE}\cdot
\frac{dE}{dT}\propto E^{-\alpha_{\rm E}}\cdot E^{-\alpha_{\rm ET}+1}\propto
T^{-1-(\alpha_{\rm E}-1)/\alpha_{\rm ET}}$, and thus $\alpha_{\rm T}=
1+(\alpha_{\rm E}-1)/\alpha_{\rm ET}$. For the peak count rate distribution,
$P\propto E/T \propto E/E^{\alpha_{\rm ET}}=E^{1-\alpha_{\rm ET}}$, then
$\alpha_{\rm P}=(\alpha_{\rm E}-\alpha_{\rm ET})/(1-\alpha_{\rm ET})$.},
which are $\alpha_{\rm T}\gtrsim 2.1$ and $1.7\lesssim\alpha_{\rm P}\lesssim2.4$,
respectively. The peak rate distribution is consistent with that of
\citet{Neilsen13}, but the duration distribution is different. As mentioned
above, the fluence-duration correlation derived in our work is also
different from that of the detected sample. Possible reasons for these
discrepancies include the incompleteness of the sample, and the profiles
of the flares. \citet{Neilsen15} showed that the flux distribution can be decomposed
into two components, a steady Poisson background and a variable
flare component. The X-ray flares excess follows a power-law
process with $\alpha_{\rm F}=1.92^{+0.03}_{-0.02}$ and the Poisson background
rate is $(5.24\pm0.08)\times10^{-3}~$cts~s$^{-1}$. The background rate
derived here $(5.90\pm0.14)\times10^{-3}~$cts~s$^{-1}$ is slightly larger than that
derived in Neilsen et al. (2015), possibly due to the different methods adopted.
Although not shown explicitly, our result of Fig.~\ref{fig:nsf_cr} gives roughly
$\alpha_{\rm F}\sim2.2$ which is consistent
with that of \citet{Neilsen15}. \citet{Wang15} also analyzed the detected
sample of \citet{Neilsen13}, with different selection of the fitting data
ranges, and gave that $\alpha_{\rm E}=1.8\pm0.6$ and $\alpha_{\rm T}=
1.9\pm0.5$\footnote{Concerning their analysis of \sgra, however, the flares sample isn't corrected
for the incompleteness,  as well as the correlation of the uncertainties among different bins of the cumulative distributions are not properly addressed in the fittings.}.
Within the large uncertainties, their results are consistent with what
we obtain.

\section{Implications on the nature of the flares}

\subsection{Confronting the Statistical Results with SOC theory}

We find that both the flare fluence distribution and the correlation
between the duration and the fluence can be described by power-laws,
which is consistent with the prediction of the {\it fractal-diffusive SOC}
theory \citep{Aschwanden11,Aschwanden12,Aschwanden14}.
The SOC theory further predicts that the power-law indices will depend
on the Euclidean space dimension $S$ of the system to produce flares.
The predicted power-law indices are $\alpha_{\rm E}^{\rm th}\approx1.5$
and $\alpha_{\rm ET}^{\rm th}\approx0.5$, respectively, for $S=3$,
the classical diffusion parameter $\beta=1$ and the mean fractal
dimension $D_{\rm S}=(1+S)/2$ (see the Appendix for more details).
These values may vary in a range for different assumed values of $\beta$
and $D_{\rm S}$. For example, $\alpha_{\rm E}$ can range from 1.4 to 1.7
for the fractal dimension $1\leq D_{\rm S=3}\leq 3$. Our results of
$\alpha_{\rm E}$ and $\alpha_{\rm ET}$ are actually in good agreement
with the theoretical expectation with $S=3$. The predicted indices of
the duration and peak rate distributions are $\alpha_{\rm T}^{\rm th}
\approx2.0$ and $\alpha_{\rm P}^{\rm th}\approx1.7$. As a comparison,
our induced values are $\alpha_{\rm T}\gtrsim 2.1$ and $1.7\lesssim\alpha_{\rm P}
\lesssim2.4$, respectively. There are potential discrepancies of these
two distributions. Since the determination of the peak rate depends on
the assumption of the flare profile, as well as the precise measurement
of the flare duration, there should be relatively large uncertainty
of the peak CR distribution. As shown in \citet{Nowak12}, the profile,
at least for the bright ones, is indeed asymmetric rather than Gaussian.
Therefore, the integral property (fluence) should be more reliably measured and
more suitable to be used to compare with the theoretical model expectation.

\subsection{Episodic ejection of plasma blobs as origin of flares?}

Two main conclusions can thus be obtained from the above analysis.
First, the power-law distributions of the fluences, durations, and their
correlation, suggest that the flares of \sgra\ can be explained in the
{\it fractal-diffusive} SOC framework. Second, the inferred space
dimension responsible for the flares is $S=3$. Both results are similar
to the solar flares, which thus implies that the X-ray flares of \sgra\
are likely driven by the similar physical mechanism as that of the solar
flares. By analogy with the coronal mass ejections and their solar flares,
\citet{Yuan09} have proposed a magnetic reconnection model
for the episodic ejections and the production of flares\footnote{The flares
and episodic ejections are physically associated with each other, both for
the Sun and the black hole accretion flow.} from the accretion flow of
\sgra. In the following, we briefly review the key points of the model.

The structure of the accretion flow of \sgra\ is quite similar to the atmosphere
of the Sun, i.e., a dense disk enveloped by a tenuous corona, as shown by the
magneto-hydrodynamic simulations \citep[e.g., Fig. 4 in][]{Yuan14}.
The magnetic loops emerge into the disk corona due to the Parker
instability. The configuration of the coronal magnetic field emerging
from the accretion flow is controlled by the convective turbulence motion
of the plasma in the disk. Since the foot points of the field lines are
anchored in the accretion flow which is turbulent and differentially
rotating, a swamp of small-scale magnetic reconnection sets in, which
redistributes the helicity and stores most of it in a flux rope.
The turbulent processes in the accretion flow thus continuously build
up magnetic stress and helicity. When a threshold is reached, e.g.,
when the current density inside the current sheet below the flux rope
is strong enough, microscopic instabilities such as the two-streaming
instability would be triggered, resulting in anomalous resistivity
and fast magnetic reconnection inside the current sheet \citep{Chen11}. The equilibrium
of the flux rope breaks down with accompanied dissipation of magnetic
energy in a catastrophic manner, powering the observed flare.
This is the so-called much-awaited SOC state which means that a small perturbation, owing
to turbulence motion in accretion flow and subsequent magnetic
reconnection, will trigger an avalanche-like chain reaction of any
size in the system once the system self-organizes to a critical state.

\section{Conclusions and Discussion}
We present a statistical analysis to the {\it Chandra} $2-8$ keV X-ray
light curve of \sgra\ from the 3 Ms XVP campaign. A Monte Carlo simulation
method is adopted to generate the model light curves taking into account
for the Possion fluctuations of both the photon counts and the number of
flares. Then we fit the CR distribution and the SF of the light curves
jointly to constrain the model parameters, via an MCMC method. We find
that the X-ray emission of \sgra\ can be well modelled by two distinct
components: a steady quiescent emission around the Bondi radius with a
CR of $(5.90\pm0.14)\times10^{-3}~\rm{cts\ s}^{-1}$, and a flaring emission with
a power-law fluence distribution $dN/dE\propto E^{-\alpha_{\rm E}}$ with
$\alpha_{\rm E}=1.65\pm0.17$. The duration-fluence correlation can
also be modelled by a power-law form $T\propto E^{\alpha_{\rm ET}}$ with
$\alpha_{\rm ET}<0.55$.

These statistical properties are consistent with the theoretical predications
of the SOC system with the Euclid spatial dimension $S=3$, same as that
for solar flares \citep{Aschwanden12,Aschwanden14}. Our analysis, therefore,
indicates that the X-ray flares of \sgra\ are possibly driven by the same
physical mechanism as that of the solar flares, i.e., magnetic reconnection \citep{Shibata11}.
This idea is further supported by the recent development of
magneto-hydrodynamic simulations which lead to the consensus that the
accretion flow in \sgra\ is enveloped by a tenuous corona above the dense
disc \citep[e.g.,][]{Yuan14}, similar to the atmosphere of the Sun.
The three dimensional geometry of the energy dissipation domain further
suggests that the X-ray flaring of \sgra\ occurs in the surface of the
accretion flow due to the less-ordered magnetic field structure embedded
in it compared with that in the relativistic astrophysical jets \citep[e.g., the GRBs;][]{WD13}.

The flares of \sgra\ have also been detected in the NIR band \citep[e.g.,][]{Genzel03,Ghez04}. \cite{DoddsEden11} and \cite{Witzel12} have analysed the NIR flare data and found that the NIR flux distribution can also be described by a power-law form. However, the power-law indices obtained, which are $-2.1\pm0.6$ \citep{DoddsEden11} and $-4.2\pm0.1$ \citep{Witzel12}, are different from each other. The cause for this discrepancy between these two results is unclear, and may be related to the subtractions of the stellar light. The former is roughly consistent with that of the X-ray flares \citep[e.g.,][as well as our results in the context of the SOC framework with $S=3$]{Neilsen15}. If this is the case, then the NIR flares are expected to have the same physical origin as the X-ray flares, which is also supported by
the results that NIR and X-ray flares occur simultaneously when there are coordinated observations at two wavebands \citep[e.g.,][]{DoddsEden09}. Otherwise, it may suggest different radiation mechanisms between the NIR and X-ray flares which lead to the differences in the fractal dimension \citep{Aschwanden11} or even different physical origins \citep{Chan15}.

By analogy with solar flares, the multi-waveband flares of \sgra\ in X-ray, probably NIR, and less prominent sub-millimeter and radio, are likely associated with the ejection of plasmoids, both of which are the radiative
manifestation of the common catastrophic phenomena \citep{Yusef06,Yusef08,Brinkerink15}. Such a result would be useful for understanding the origins of the flares and episodic jets in various black holes in general.

Mineshige et al. (1994b; see also \citealt{Negoro95,Takeuchi95}) studied the statistical properties of the X-ray fluctuations in Cygnus X-1, a black hole X-ray binary. They found that the X-ray fluctuations showed much steeper (or even exponential) distributions compared with those of solar flares. They argued that the fluctuations should also follow an SOC process. A mass diffusion process was invoked in their works to explain the discrepancy of the distributions. But the X-ray fluctuations they discussed likely have a different physical origin compared with what we have discussed in the present paper. Thus the trigger mechanism of SOC is also different. 

\acknowledgments
We thank all the members of the \sgra\ {\it Chandra} XVP collaboration (with PIs Fred Baganoff, Mike Nowak and Sera Markoff)\footnote{http://www.sgra-star.com/collaboration-members}, and we are immensely grateful to {\it Chandra} Mission Planning for their support during our 2012 campaign. We thank F.Y. Wang, G. Witzel, S. Gillessen, D. Haggard, and S. Nayakshin for beneficial discussions. We would like to thank an anonymous referee for the useful suggestions. YPL acknowledges Graduate School of Xiamen University for the short term exchange program in University of Massachusetts Amherst. FY was supported in part by the National Basic Research Program of China (973 Program, grant 2014CB845800), the Strategic Priority Research Program ``The Emergence of Cosmological Structures" of CAS (grant XDB09000000), and NSF of China (grant 11133005). PFC was supported in part by the National Basic Research Program of China (973 Program, grant 2011CB811402) and the NSF of China (grant 11025314).
%This work was supported by the National Basic Research Program
%of China (973 Program, grants 2014CB845800 and 2011CB811402), the NSFC
%(grants 11133005, 11121062 and 11025314), and the Strategic Priority
%Research Program ``The Emergence of Cosmological Structures" of CAS
%(Grant XDB09000000).

%\clearpage

%\begin{minipage}{1.0\textwidth}

%\end{minipage}

\begin{appendix}

%\pagebreak
\section{Self-Organized Criticality System}

The theoretical concept of SOC, proposed first by \citet{Katz86} and \citet{BTW} independently, describes
a class of nonlinear dissipative dynamical systems which are slowly and
continuously driven toward a critical value of an instability threshold,
producing scale-free, fractal-diffusive, and intermittent avalanches
\citep{Charbonneau01,Aschwanden11}. The classical example of the SOC
system is a sandpile \citep{BTW}. If one continuously drops the sand
grains (driving force) to the same place, a conical sandpile will grow
in a steady way until the surface shape reaches a critical slope, beyond
which further addition of sand rapidly leads to catastrophic
avalanches of the sandpile. The critical slope is primarily determined
by the friction threshold between adjacent grains. Such a threshold is
crucial since it allows the existence of multiple metastable states
across the system. The continuous slow addition of sand will produce
small or large avalanches with sizes independent of the input rate of
sand. The critical behavior of the sandpile appears naturally as a
consequence of the slow addtion of sand without fine tuning of the
addition rate. It is in this sense that the system is self-organized
to criticality \citep{Charbonneau01,Aschwanden11}. The continuous energy
input (addition of sand) and nonlinear energy dissipation (avalanches due
to the complicated interactions between the colliding sand grains) are
two key points to determine whether a system is in an SOC state.

A universal property of the SOC system is that there is no preferred
scale of the release of energy. The scale-free power-law distributions
of various event parameters, which actually become the hallmarks of SOC
systems, are expected. The ubiquitousness of power-law distributions in
a wide range of physical systems, e.g., earthquakes, cloud formation,
solar flares, and widely in astrophysics, suggests that SOC is a common
law of the nature. The SOC theory has been widely applied in geophysics
\citep{Turcotte99}, solar physics \citep[e.g.,][]{Charbonneau01}, and astrophysics
\citep[e.g.,][]{Mineshige94a,Kawaguchi00,Aschwanden11,Kunjaya11}.

The slope of the power-law form depends on how the subsystem self-organizes
to the critical state. An analytic description was provided based on a
{\it fractal-diffusive} avalanche model \citep{Aschwanden12}. The SOC
state ensures that the entire system is close to the instability threshold,
and the avalanches can develop in any direction once triggered. The
propagation of the unstable nodes can thus be modelled with a random
walk in an $S$-dimensional space \citep{Aschwanden12}. A statistical
correlation between the spatial length scale $L$ and time duration of
the avalanche $T$ for a diffusive random walk is predicted to be
$L\propto{T}^{\beta/2}$, where $\beta$ is a diffusion parameter ($\beta=1$
corresponds to the classical diffusion). With the scale-free probability
conjection $dN/dL\propto {L}^{-\rm S}$, which means that the critical
states are homegeneously distributed across the entire system, the
occurrence frequency distributions of flare duration, flux, peak flux,
and total energy can be obtained in the {\it fractal-diffusive} avalanche
model \citep{Aschwanden12}. The indices of the total energy (fluence)
$\alpha_{\rm E}^{\rm th}$, duration $\alpha_{\rm T}^{\rm th}$, peak
luminosity (or peak CR in this work) $\alpha_{\rm P}^{\rm th}$, and flux
(or CR) $\alpha_{\rm F}^{\rm th}$ distributions are derived as
\citep{Aschwanden12,Aschwanden14}
\begin{eqnarray}
  \alpha_{\rm E}^{\rm th}&=&1+\frac{S-1}{D_{\rm S}+2/\beta}, \label{NE} \\
  \alpha_{\rm T}^{\rm th}&=&1+\frac{(S-1)\beta}{2}, \label{NT}\\
  \alpha_{\rm P}^{\rm th}&=&1+\frac{S-1}{S}, \label{NP} \\
  \alpha_{\rm F}^{\rm th}&=&1+\frac{S-1}{D_{\rm S}}, \label{NF}
\end{eqnarray}
where $S$ is the Euclidean space dimension, $D_{\rm S}$ is the fractal
{\it Hausdorff} dimension which lies between 1 and $S$. The {\it
fractal-diffusive} SOC theory also predicts the scaling relations between
various flare parameters. The correlation slope between the total energy
$E$ and duration $T$ is \citep{Aschwanden12,Aschwanden14}
\begin{equation}\label{ET}
  \alpha_{\rm ET}^{\rm th}=\frac{2}{D_{\rm S}\beta+2}.
\end{equation}

For the classical diffusion $(\beta=1)$ and an estimated mean fractal
dimension of $D_{\rm S}\approx(S+1)/2$, we have $\alpha_{\rm E}^{\rm th}
=1.5$, $\alpha_{\rm T}^{\rm th}=2.0$, $\alpha_{\rm P}^{\rm th}=1.7$,
$\alpha_{\rm F}^{\rm th}=2.0$, and $\alpha_{\rm ET}^{\rm th}=0.5$ for $S=3$.

\end{appendix}

\end{document}